%% file: main.tex
\documentclass[10pt,conference]{IEEEtran}
\usepackage[square, comma, numbers]{natbib}
\IEEEoverridecommandlockouts
\pdfoutput=1

\AtBeginDocument{%
  \providecommand\BibTeX{{%
    \normalfont B\kern-0.5em{\scshape i\kern-0.25em b}\kern-0.8em\TeX}}}
\usepackage{comment}
\usepackage{tabularx}
\usepackage{makecell}
\usepackage{threeparttable}
\usepackage{amsmath}

\usepackage{amsfonts}
\usepackage{amssymb}
\usepackage{mathrsfs}
\usepackage[pdftex]{graphicx}
\usepackage{multirow}
\usepackage{color, colortbl}
\usepackage[inline]{enumitem}
\usepackage[dvipsnames]{xcolor}
\usepackage{pgfplots}
\usepackage{tikz}
\usepackage{tikzscale}
\usepackage{soul}
\usepackage{enumitem}
\usepackage{etoolbox}
\usepackage{xcolor}
\usepackage{tabulary}
\usepackage{subcaption} 
\usepackage[colorinlistoftodos]{todonotes}
\usepackage{algorithm, algorithmic}
\usepackage[normalem]{ulem}
\usepackage{caption}
\usepackage[compact]{titlesec}
\usepackage{xcolor}
\usepackage{todonotes}
\usepackage{url}
\usepackage{booktabs}

\pgfplotsset{
  every axis plot/.append style={line width=1.2pt},
}

\newcommand{\cf}{\textit{cf.}, }
\newcommand{\eg}{\textit{e.g.}, }
\newcommand{\ie}{\textit{i.e.}, }

\newcommand{\fig}[1]{Fig.~\ref{#1}}

\newcommand{\sect}[1]{Section~\ref{#1}}

\pgfplotsset{compat=1.14}
\usepackage{newtxmath}
\usepackage{bbm}

\usetikzlibrary{arrows.meta, fit, backgrounds, shapes.geometric}
\tikzset{%
  >={Latex[width=2mm,length=2mm]},
         base/.style = {rectangle, rounded corners, draw=black,
                           minimum width=3cm, minimum height=1cm,
                           text centered, font=\rmfamily},
         terminal/.style = {base, circle, minimum size=1.5cm,font=\rmfamily}
}

\begin{document}
\bstctlcite{IEEEexample:BSTcontrol}
\title{MicroOpt: Model-driven Slice Resource Optimization in 5G and Beyond Networks}

\author{\IEEEauthorblockN{ Muhammad Sulaiman\IEEEauthorrefmark{1}, Mahdieh Ahmadi\IEEEauthorrefmark{1}, Bo Sun\IEEEauthorrefmark{1}, Niloy Saha\IEEEauthorrefmark{1}, Mohammad A. Salahuddin\IEEEauthorrefmark{1},\\Raouf Boutaba\IEEEauthorrefmark{1}, Aladdin Saleh\IEEEauthorrefmark{2} \\}
\IEEEauthorblockA{\{m4sulaim, b24sun, m4ahmadi, n6saha, mohammad.salahuddin, rboutaba\}@uwaterloo.ca, aladdin.saleh@rci.rogers.com,}
\IEEEauthorblockA{\IEEEauthorrefmark{1}University of Waterloo, 
\IEEEauthorrefmark{2}Rogers Communications Canada, Inc.} 
}

\maketitle
\input{abstract}
\begin{IEEEkeywords}
5G, Network Slicing, Dynamic Resource Scaling, Machine Learning, Quality of Service
\end{IEEEkeywords}

\input{introduction}

\input{related}

\input{design}

\input{implementation}

\input{experiments}

\input{conclusion}

\section*{Acknowledgement}
This work was supported in part by Rogers Communications Canada Inc. and in part by a Mitacs Accelerate Grant.

\newpage\small
\bibliographystyle{IEEEtranN}
\bibliography{IEEEabrv,main}

\end{document}

%% file: abstract.tex
\begin{abstract}

A pivotal attribute of 5G networks is their capability to cater to diverse application requirements. This is achieved by creating logically isolated virtual networks, or slices, with distinct service level agreements (SLAs) tailored to specific use cases. However, efficiently allocating resources to maintain slice SLA is challenging due to varying traffic and QoS requirements. Traditional peak traffic-based resource allocation leads to over-provisioning, as actual traffic rarely peaks. Additionally, the complex relationship between resource allocation and QoS in end-to-end slices spanning different network segments makes conventional optimization techniques impractical.  Existing approaches in this domain use network models or simulations and various optimization methods but struggle with optimality, tractability, and generalizability across different slice types. In this paper, we propose MicroOpt, a novel framework that leverages a differentiable neural network-based slice model with gradient descent for resource optimization and Lagrangian decomposition for QoS constraint satisfaction. We evaluate MicroOpt against two state-of-the-art approaches using an open-source 5G testbed with real-world traffic traces. Our results demonstrate up to 21.9\% improvement in resource allocation compared to these approaches across various scenarios, including different QoS thresholds and dynamic slice traffic. 

\end{abstract}

%% file: introduction.tex
\section{Introduction}\label{sec:introduction}
Network slicing empowers 5G networks to accommodate applications and services with diverse Quality of Service (QoS) requirements \citep{foukas2017network} such as ultra low latency or high throughput. Facilitated by Software Defined Networking (SDN) and Network Function Virtualization (NFV), network slicing allows the infrastructure providers (InPs) to virtualize their physical network resources and create isolated virtual networks on top of a shared infrastructure \cite{yousaf2017nfv}. For example, enhanced mobile broadband (eMBB) slices can cater to high-throughput applications such as 4K video streaming, while ultra-reliable low-latency communication (URLLC) slices can support latency-sensitive applications such as remote surgery \cite{alsenwi2021intelligent, ali2021urllc}. However, network slicing also brings forth the challenge of effectively managing resources in a complex and dynamic environment. Each slice is associated with a service level agreement (SLA) specifying the peak traffic and minimum QoS requirements of slice users (\ie slice tenants). 

To ensure QoS, the InP can allocate isolated resources to each slice based on its peak traffic. However, this approach often leads to over-provisioning as the actual slice traffic may exhibit fluctuations over time and rarely reach its peak \cite{traffic_dataset}, resulting in the under-utilization of resources. Moreover, SLAs can be dynamic and subject to change based on various factors, including the number of users, slice location, and time of day. For example, a smart healthcare application requires a URLLC slice with full isolation and extremely low latency during surgery \cite{habibi2018structure, qureshi2020service} which can be adjusted post-surgery to align with updated service expectations.
This highlights the need for a more adaptable approach to resource management that is capable of accommodating changes in traffic patterns and SLA requirements.
To achieve this, the InP needs to maintain QoS degradation under a specific threshold by predicting future traffic patterns and dynamically allocating resources.  This problem is known as predictive resource allocation or dynamic resource scaling (DRS) of network slices. 

 

To achieve efficient resource utilization while meeting SLA commitments, it is crucial to carefully examine the relationship between resource allocations and QoS metrics. However, QoS metrics for an end-to-end slice, such as throughput, latency, and reliability, rely on various resource types across multiple network segments, including the Radio Access Network (RAN), transport network, and core network. Traditional network models such as queues, often fail to  capture these complexities accurately, especially considering the highly dynamic and mobile nature of 5G networks \cite{liu2021constraint}. This adds to the complexity of DRS.

Existing approaches for resource scaling typically involve traffic forecasting \cite{li2020dra} for a given slice, followed by utilizing simulations \cite{liu2021constraint,liu2022atlas} or Machine Learning (ML)  \cite{noms23} to learn the network model and determine the optimal resource allocation. Simulation-based methods, such as using packet-level simulators (\eg ns-3) \cite{liu2022atlas} or queue-based simulators \cite{liu2021constraint}, are computationally expensive \cite{ferriol2023routenet}, and may not accurately represent the network. Additionally, some of these works assume the slice traffic to be constant \cite{liu2022atlas}.

ML-based approaches model the entire network as a single entity using a neural network and have shown promising performance \cite{ferriol2023routenet}. However, existing works that follow this approach adopt a simple regression-based model, which fails to capture the complexities of an end-to-end network \cite{noms23}. Furthermore, due to its lack of a differentiable network model, it cannot take advantage of efficient gradient-based optimization techniques. Instead, it is constrained to employ an inefficient offline optimization method relying on constrained Reinforcement Learning (RL). Consequently, it may yield sub-optimal results when the network conditions deviate from the training environment. On the other hand, complete model-free approaches that aim to learn resource allocation directly in an online fashion can impact SLAs and require longer training times \cite{liu2021onslicing}.

In this paper, we introduce MicroOpt, a novel framework that combines the power of ML with continuous optimization for dynamic resource scaling of network slices. Our proposed approach leverages a neural network to estimate QoS metrics and incorporates optimization techniques for efficiently scaling slice resources. By employing the reparameterization trick \cite{vae_paper}, which is a commonly used technique in probabilistic Deep Learning (DL) models, we can iteratively refine slice resource allocation to minimize resource usage while meeting SLA requirements. The reparameterization trick ensures the differentiability of the slice model, enabling continuous optimization through gradient descent. The major contributions of this work are: \begin{itemize}[leftmargin=*]
    \item We introduce MicroOpt, a novel framework for dynamic resource scaling of 5G slices. The proposed framework is capable of accommodating time-varying traffic, dynamic QoS and QoS degradation thresholds, without the need for retraining.
    \item We present a deep neural network (DNN)-based model of 5G slices that utilizes the reparameterization trick to ensure that gradients of QoS degradation constraints, with respect to resource allocations, can be computed analytically. The model is evaluated using metrics such as mean squared error, mean absolute error, and log probability loss on previously unseen inputs. 
    \item We propose a primal-dual optimization algorithm that leverages the analytical differentiability of the slice model to enable efficient resource allocation optimization using gradient descent. Our algorithm employs an inner loop that leverages a relaxed differentiable Lagrangian function for optimizing resource allocation, whereas, in the outer loop, strict definitions are utilized to enforce QoS degradation constraints.
    \item We compare MicroOpt against two state-of-the-art baselines \cite{liu2022atlas, icnp_older}, highlighting their limitations. Additionally, we evaluate variations of our framework that incorporate the optimization techniques from these baseline works, \ie Bayesian Optimization (BO) and Constrained Deep Reinforcement Learning (CDRL). 
    \item We validate the performance of the proposed framework using a 5G testbed, with real-world traffic traces for multiple QoS, and QoS degradation thresholds. 
\end{itemize}

The structure of the paper is as follows. In \sect{sec:bg_related}, we provide a comprehensive review of related works encompassing both network modeling and resource allocation. \sect{sec:statement} formally defines the problem, while \sect{sec:sol} presents our proposed solution. \sect{sec:implementation} details our testbed and evaluation setup, and \sect{sec:experiments} presents the evaluation results. We conclude in \sect{sec:conclusion} and investigate future research directions.



%% file: related.tex
\section{Related Works}\label{sec:bg_related}

Predictive resource allocation is the process of allocating resources in anticipation of future user demand. This approach can assist in maintaining SLAs with minimum over-provisioning, even when demand spikes. 
The quintessential components of predictive resource allocation include: 
\begin{enumerate*}[label=(\roman*)]
    \item a future traffic predictor\label{modules:1}, \item a network model\label{modules:2}, and \item a resource allocation algorithm\label{modules:3}.
\end{enumerate*}
As mobile traffic prediction has been widely researched \cite{joshi2015review}, here we will focus on \ref{modules:2} and \ref{modules:3}. The network model is used to estimate QoS based on predicted demand and allocated resources, and 
the resource allocation algorithm aims to satisfy SLA constraints while minimizing the allocated resources. 
It should be noted that these modules may not be separately identifiable in all related works. For example, numerous works in the literature have assumed that resource demand can be readily derived based on traffic \cite{kasgari2018stochastic, salvat2018overbooking}, which precludes the necessity of \ref{modules:2}. Similarly, works that use RL for end-to-end resource allocation implicitly assume that RL agent learns to predict the traffic pattern in addition to the corresponding resource requirements \cite{liu2021constraint, LiWeighted}.

\subsection{Slice Modeling}
The effectiveness of resource allocation algorithms is highly dependent on slice models, which correlate allocated resources with QoS distribution based on predicted demand and slice configuration. However, such models may not exist for end-to-end slices under various traffic distributions or network configurations. Therefore, network simulators \cite{liu2021constraint, liu2022atlas} and ML-based estimators \cite{noms23} are commonly employed to address this challenge.

Conventional network simulators, such as ns-3, are packet-level and time-intensive \cite{ferriol2023routenet, yang2022deepqueuenet}, which limits their application for online resource allocation and even for offline training of RL policies \cite{iacoboaiea2022design}. 
Moreover, simulators may not accurately mimic real-world scenarios, especially in wireless domains \cite{liu2022atlas}. 
Bayesian optimization has been utilized by \citet{liu2022atlas} to identify the optimal ns-3 parameters and reduce the disparity between simulated and real-world conditions. However, this process needs to be repeated for each minor alteration in the network and SLA metrics.

ML-driven approaches model the entire network as a neural network, which can be trained to estimate end-to-end QoS metrics using traffic traces. The constraints of this approach include limited visibility at the packet-level and lack of generalizability across diverse network settings. Recently, graph neural networks \cite{ferriol2023routenet} and a combination of simulation and DNN models \cite{yang2022deepqueuenet} have been used to alleviate these concerns in the context of  transport networks. However, the transport network corresponds to only a single portion of an end-to-end slice, and usually handles aggregated traffic. \citet{noms23} use a simple regression model to estimate the QoS distribution of a single slice with only one resource type. However, since their QoS degradation derivation requires non-deterministic operations on the network model's output, they cannot leverage gradient backpropagation through the network model for optimization.

\subsection{Resource Allocation Algorithm}
\textbf{Machine Learning.} ML-based optimization approaches have been successfully applied in various resource allocation \cite{sulaiman2022coordinated} and scheduling problems \cite{sciancalepore2019rl}. In the context of predictive resource allocation, \citet{bega2019deepcog} use an encoder-decoder model to predict the amount of resources needed to minimize the aggregated cost of resource over-provisioning and SLA violations due to customer churn. However, the aggregate cost function does not guarantee SLAs.  

Recently, a number of works have used constrained deep reinforcement learning (DRL) techniques to learn the optimal resource allocation under average SLA constraints \cite{noms23, liu2021constraint, liu2021onslicing}. These approaches require retraining for each minor change in SLA and may not  generalize to real traffic patterns that are unseen during training. \citet{liu2021constraint} utilize queue-based simulated environment to  train the resource allocation policy offline. This approach may encounter issues stemming from the discrepancy between the simulated environment and the real-world network. Additionally,  \citet{liu2021onslicing} adopt a fully model-free approach by training the policy online. However, this method seems impractical for implementation in a production network due to long convergence time.

\citet{noms23} address some of these drawbacks by using a simple regression-based model for QoS estimation and a risk-aware CaDRL agent trained offline over randomized traffic to satisfy the SLA constraint. 
However, since RL algorithms do not leverage the network model's gradient for policy optimization, they have to rely on inefficient random exploration. In soft actor-critic algorithms adopted in this work, this is achieved by learning a critic network to estimate the state-action value of different actions and a policy network that maximizes the expected cumulative reward. However, as shown in \cite{noms23}, even though the offline learned policy generalizes to different online scenarios, it achieves sub-optimal results compared to approaches trained with advanced knowledge of the online scenarios.

\textbf{Optimization.} Many works in the literature assume that the amount of required resources can be readily determined based on the SLA requirement, and utilize optimization techniques to improve performance in scenarios with multiple slices and resource contention \cite{kasgari2018stochastic, salvat2018overbooking}. 
Although this may hold for Physical Resource Block (PRB) allocation in a single base station, it cannot be generalized to an end-to-end slice that requires the allocation of various resource types across different network segments, such as the bandwidth on the transport network, and compute resources for running virtualized network functions and edge applications.

\citet{liu2022atlas} employ Bayesian optimization (BO) to minimize the resource consumption of a single slice while satisfying its SLA. The authors relax the constraint using the Lagrangian method and approximate QoS using a Bayesian neural network. 
However, the Bayesian neural network approximates the probability of violating the QoS rather than approximating the QoS itself. Moreover, for each slice and traffic value, QoS is approximated for a limited set of resource allocations that have a higher chance of being the solution. 
Therefore, if there are changes in traffic or SLA, the neural network needs to query new points in the domain. Additionally, the model assumes a constant level of traffic throughout the entire configuration interval (\ie 1-2 hours), which is not the case in practice. This assumption compels to consider the worst-case traffic value within each interval to satisfy the SLA, which leads to a sub-optimal solution. Moreover, even with an assumption of constant traffic, the convergence time of this method is in the order of hours, which can span a significant portion of each reconfiguration interval.

In this work, we address two main drawbacks identified in the current literature. First, large QoS querying times---network simulators, such as ns-3 \cite{liu2022atlas}, and queue-based network models \cite{liu2021constraint, liu2021onslicing}, are computationally intensive and require significant time to compute the QoS for a given resource allocation. Second, ineffective exploration---optimization methods like RL, which use $\epsilon$-greedy algorithms for solution space exploration, require numerous interactions with the network and often leading to sub-optimal solutions. While modern network virtualization solutions, such as Kubernetes and ONOS, enable rapid reconfiguration of resource allocation, these current approaches fail to fully leverage this capability due to their inherent inefficiencies.

We address the first challenge by employing a DNN-based slice model to predict QoS, enabling fast queries on the order of milliseconds once trained. To tackle the second challenge, we utilize the reparameterization trick and gradient-based optimization, which leverages gradient information to adjust the search direction. This approach efficiently explores the parameter space, resulting in a more optimal solution, and faster convergence.

%% file: design.tex
\section{Problem Statement} \label{sec:statement}
The process of establishing an end-to-end network slice involves the allocation of resources across various network segments. 
Reconfiguration of these resources occurs at specific time intervals referred to as reconfiguration intervals. We use $i$ to denote the $i^{th}$ reconfiguration interval. 
The duration of these intervals, represented by $\tau_i$, depends on various factors, including data collection delays \cite{liu2021onslicing}, constraints imposed by the substrate network (\eg the use of legacy virtual infrastructure manager) \cite{DTI1}, or the time necessary for accurate traffic forecasting \cite{DTI2}, to name a few. 
Typically, the duration of a reconfiguration interval can range from several minutes \cite{liu2021onslicing} to hours \cite{liu2022atlas}. Additionally, $\tau_i$ can be dynamically adjusted to account for unexpected situations in the real world. We denote the set of operational network slices as ${S}$, where each slice $s \in S$ is associated with a resource allocation vector $\boldsymbol{r}_{i}^s = [r_{i}^{s, 1}, \cdots, r_{i}^{s, K}] \in \mathbb{R}_{\geq 0}^{K}$, comprising $K$ resources.

The slice traffic, \ie the number of users connected to a slice, may exhibit variations during reconfiguration intervals \cite{traffic_dataset}. To represent the traffic time series within interval $i$, we employ the notation $\boldsymbol{x}_{i}^s = [{x}^s_{i}(1), \cdots, {x}^s_{i}(\tau_i)] \in \mathbb{R}_{\geq 0}^{\tau_i}$. 
The traffic is measured in users/s and the average QoS experienced by these users is determined by the resources allocated to the slice. We represent the average QoS for slice $s$ in interval $i$ as $\boldsymbol{q}_{i}^s = [q_{i}^s(1), \cdots, q_{i}^s(\tau_i)] \in \mathbb{R}_{\geq 0}^{\tau_i}$. 
For each time slot, if the QoS fails to surpass the predetermined QoS threshold $q^s_{\textit{thresh}}$, it results in QoS degradation. Under the assumption of fair resource allocation and equal average QoS experienced by each user, the average QoS degradation for slice $s$ during the reconfiguration interval $i$ is defined as:

 \begin{equation} \label{beta}
     \begin{aligned}
         \beta^s_{i} = \frac{\sum_{t=1}^{\tau_i} {{{x}^s_{i} (t)} \cdot {\mathbbm{1}_{[q^s_{i}(t) \leq q^s_{\textit{thresh}}]}}}}{ {\sum_{t=1}^{\tau_i}{x}^s_{i}(t)}}.
     \end{aligned}
 \end{equation}

The objective is to minimize the normalized resource allocation to network slices within each reconfiguration interval $i$, while respecting the QoS degradation constraints~\cite{noms23, liu2022atlas, DTI2}. This can be formulated as:

\begin{equation} \label{obj}
\begin{aligned}
\min_{\mathbf{r^s_{i}} \geq 0 } \quad & \sum_{s \in \mathbf{S}} \boldsymbol{\eta}^\intercal  \mathbf{r}^s_{i},\\
\textrm{s.t.} \quad & \mathbbm{E} \left( \beta^s_{i}\right) \leq \beta^s_{\textit{thresh}}, \;\;\;\forall s \in S\,\\
\quad & \sum_{s \in S} \mathbf{r}_{i}^s \leq {\mathbf{R}}.
\end{aligned}
\end{equation}

where $\boldsymbol{\eta} \in \mathbb{R}_{>0}^{K}$ is the normalization vector, and can also be used to represent the priority or the relative cost associated with different resources, $\mathbbm{E}$ represents the expectation operator with respect to the QoS distribution, and vector $\mathbf{R}=[R^1, \cdots, R^K] \in \mathbb{R}_{\geq 0}^{K}$ represents the capacities of different resources. The aforementioned problem pertains to a black-box continuous optimization problem due to the unknown distribution of QoS, $q_{i}^s(t)$. 
In the following section, we will introduce the MicroOpt framework designed to tackle this problem.

\section{MicroOpt Framework}\label{sec:sol}

\begin{figure}[ht!]
\captionsetup{justification=centering}
  \centering
  \begin{minipage}{\columnwidth}
  \centering
  \includegraphics[width=\linewidth]{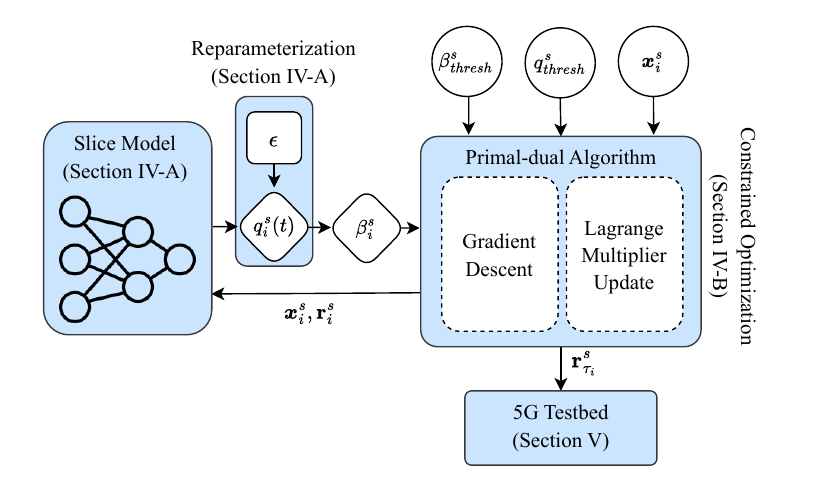}
    \caption{MicroOpt framework overview}
    \label{fig:system_model}
  \end{minipage}
\end{figure}

We propose MicroOpt, shown in \fig{fig:system_model}, a framework for continuous optimization of network slice resources, that solves the constrained optimization problem in \eqref{obj} by addressing three key aspects: 
(i) predicting slice traffic, (ii) obtaining the slice model function, and (iii) solving the constrained optimization problem. We assume that slice traffic exhibits regular patterns \cite{traffic_dataset} and can be predicted using off-the-shelf time-series forecasting techniques (\eg  \cite{neural_prophet}). To model the slice, we employ a DNN-based approach, which has been successfully employed in the network digital twin space \cite{ferriol2023routenet, yang2022deepqueuenet}. 
This approach leverages the expressive power of neural networks to capture the complex relationships and dependencies within the slice, and allows for analytical gradient calculation through the reparameterization trick. Finally, for tackling the constrained optimization problem, we propose a primal-dual optimization algorithm that capitalizes on the differentiability of the neural network for efficient online optimization.


\subsection{Slice Model} \label{sec:nn_model}
\begin{figure}[ht!]
\captionsetup{justification=centering}
  \centering
  \begin{minipage}{\columnwidth}
  \centering
  \includegraphics[width=\linewidth]{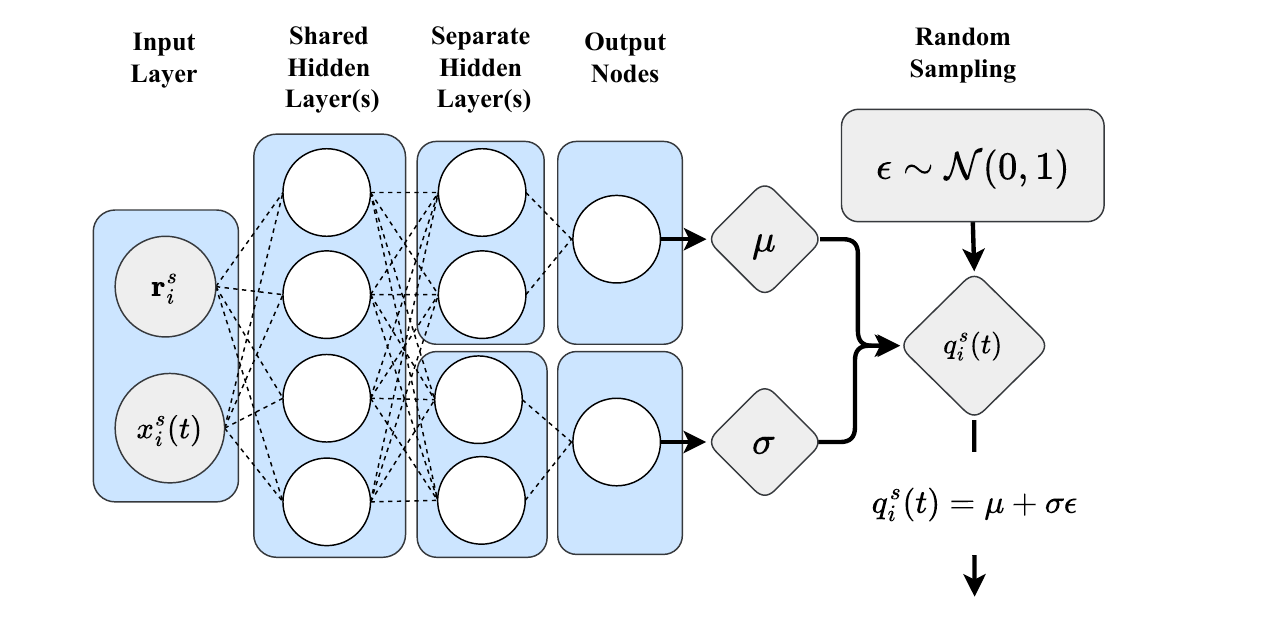}
    \caption{Slice model}
    \label{fig:net_model}
  \end{minipage}
\end{figure}

Slice modeling encompasses the acquisition of the function $f^s_{\textit{QoS}}(x_{i}^s(t), \mathbf{r}_{i}^s)$, that captures the relationship among resource allocation $\mathbf{r}_{i}^s$, slice traffic $x_{i}^s(t)$, and QoS distribution. The QoS sampled from this distribution, \ie $q_{i}^s(t) \sim f^s_{\textit{QoS}}(x_{i}^s(t), \mathbf{r}_{i}^s)$, can be used to calculate the QoS degradation $\beta^s_{i}$ using \eqref{beta}. Finally, the estimated $\beta^s_{i}$ is used for solving the constrained optimization problem in \eqref{obj}. In this work, we propose using a DNN model to learn $f^s_{\textit{QoS}}(x_{i}^s(t), \mathbf{r}_{i}^s)$ using a QoS dataset encompassing various resource allocations ($\mathbf{r}_{i}^s$) and traffic ($\boldsymbol{x}_{i}^s$).
Unlike mathematical models or queues, a DNN-based model can easily handle heterogeneous types of resources and predict different QoS metrics. Additionally, the complexity of this approach does not depend on the traffic volume, which is the case with packet-level simulators.

We assume the QoS normally distributed for the remainder of this paper, with the network model designed to predict the parameters of the QoS distribution. We choose the normal distribution as it proves to be sufficient for effectively modeling the data in our case. However, it is important to note that the proposed slice model can be extended to incorporate mixture density networks (MDNs), which have the ability to represent arbitrarily complex distributions \cite{mdn}. The architecture of the slice model is shown in \fig{fig:net_model}. The inputs to the  model consist of the slice traffic ${x}_{i}^s(t)$ and resource allocation $\mathbf{r}_{i}^s$. These inputs are connected to a set of shared hidden layers, followed by separate hidden layers dedicated to each Gaussian distribution parameter. As a result, the model outputs the Gaussian distribution parameters $\mu$ and $\sigma$ associated with the predicted QoS distribution. The probability density function of $q_{i}^s(t)$ under this distribution can be written as:
\begin{equation}
\begin{aligned}
p(q_{i}^s(t)\,|\,{x}_{i}^s(t), \mathbf{r}_{i}^s) = \frac{1}{\sqrt{2\pi}\sigma} \exp\left(-\frac{\left(q_{i}^s(t) - \mu\right)^2}{2\sigma^2}\right).
\end{aligned}
\end{equation}
Finally, the loss for the model is computed as:
\begin{equation}\label{eq:lqos}
\begin{aligned}
L_{QoS} = -\frac{1}{B} \sum_{j=1}^{B} \log p(q_{i, j}^{s}(t) | \mathbf{r}_{i, j}^{s}, {x}_{i, j}^{s}(t)),
\end{aligned}
\end{equation}
where $B$ is the batch size and the subscript $j$ represents the $j^{th}$ sample in the batch. This loss function calculates the negative log-likelihood of the QoS $q_{i, j}^{s}(t)$ under the normal distribution $\mathcal N(\mu, \sigma)$ generated by the model for the input $({x}_{i, j}^{s}(t),\mathbf{r}_{i, j}^{s})$. Once trained, this model can be used to sample the QoS from the predicted distribution. 
The detailed parameter settings and slice model results are given in Section~\ref{sec:experiments}.

The drawback of na{\"i}vely sampling QoS from this distribution is that any subsequent optimization algorithm that relies on the sampled QoS cannot leverage its gradients. This is because random sampling from a distribution involves non-deterministic operations on the model outputs, which happen to be non-differentiable. To address this, we propose using the reparameterization trick, which has been employed in the ML literature \cite{vae_paper}, and can also be used with other probability distributions including MDNs \cite{mdn_reparam}. For this purpose, the QoS sampling is reformulated as follows: 
 \begin{equation}
\begin{aligned}
q_{i}^{s}(t) = \mu + \sigma \epsilon,
\end{aligned}
\end{equation}
where $\epsilon$ is a random sample from a standard normal distribution $\mathcal N(0, 1)$ that does not depend on the input $({x}_{i}^s(t), \mathbf{r}_{i}^s)$. The reparameterization trick not only allows for more efficient gradient calculation, it also allows the use the existing automatic differentiation frameworks such as PyTorch \cite{torch} for easy implementation. Finally, after computing gradients through this operation and use them in subsequent optimization described in the following subsection.  

 \subsection{Constrained Optimization} \label{sec:motivation} \label{sec:opt}


We start by converting the constrained problem into an unconstrained one by using dual Lagrangian relaxation. The Lagrangian is defined as follows:

\begin{equation} \label{lagrangian}
\begin{aligned}
\mathcal{L}(\mathbf{r}_{i}^{s}, \boldsymbol{\lambda}, \boldsymbol{\mu}) = \sum_{s \in {S}} \boldsymbol{\eta}^\intercal  \mathbf{r}^s_{i} + \sum_{s \in S} \lambda_s \left( \mathbbm{E} \left( \beta^s_{i}\right) - \beta^s_{\textit{thresh}} \right) \\+\,\, \sum_{k=1}^{K} \mu_k \left( \sum_{s \in S} {r}_{i}^{s,k} - R^k \right),
\end{aligned}
\end{equation}
where $\boldsymbol{\lambda} = [\lambda_1, \lambda_2, \ldots, \lambda_{|S|}]$ denotes the vector of Lagrange multipliers for the QoS degradation constraints, and $\boldsymbol{\mu} = [\mu_1, \mu_2, \ldots, \mu_K]$ denotes the vector of Lagrange multipliers for the resource constraint. Based on this Lagrangian formulation, the dual problem can be written as:

\begin{equation}
\begin{aligned}
\max_{\boldsymbol{\lambda}, \boldsymbol{\mu}} \quad & g(\boldsymbol{\lambda}, \boldsymbol{\mu}) \
\text{subject to} \quad & \boldsymbol{\lambda} \geq 0, \quad \boldsymbol{\mu} \geq 0,
\end{aligned}
\end{equation}
where the dual function $g(\boldsymbol{\lambda}, \boldsymbol{\mu})$ is defined as:
\begin{equation}
\begin{aligned}
g(\boldsymbol{\lambda}, \boldsymbol{\mu}) = \inf_{\mathbf{r}_{i}^{s}} \mathcal{L}(\mathbf{r}_{i}^{s}, \boldsymbol{\lambda}, \boldsymbol{\mu}).
\end{aligned}
\end{equation}
The above dual problem could be solved iteratively using primal-dual updates with gradient-based methods \cite{boyd2004convex}, if it were differentiable with respect to both primal and dual variables. This is because gradient-based methods rely on the ability to compute the gradients of the objective function and the constraints with respect to relevant variables. However, the computation of the QoS degradation $\beta^s_{i}$ using \eqref{beta} involves an indicator function ${\mathbbm{1}{[q^s_{i}(t) \leq q^s_{\textit{thresh}}]}}$, which is piece-wise constant and has a gradient of zero almost everywhere. This poses a challenge for gradient-based optimization algorithms that rely on gradient calculations for parameter updates \cite{cotter2019two}. 

To address this challenge, we introduce a surrogate QoS degradation function $\hat{\beta}^s_{i}$ that replaces the indicator function in \eqref{beta} with a Sigmoid function $\sigma(\rho*(q^s_{i}(t) - q^s_{\textit{thresh}}))$, where  $\rho$ is a hyperparameter that controls the sharpness of the curve. 
The Sigmoid function is a smooth differentiable function and allows the use of gradient-based optimization methods, while still approximating the behavior of the indicator function.
We denote the surrogate Lagrangian function, which incorporates the surrogate QoS degradation function, as $\mathcal{\hat{L}}(\mathbf{r}_{i}^{s}, \boldsymbol{\lambda}, \boldsymbol{\mu})$. With this formulation, we can apply analytical gradient optimization techniques to optimize the resource allocation, while the solution's feasibility is ensured using the strict definition of QoS degradation. It is worth noting that if the optimization problem is non-convex, gradient-based approaches may not always converge to the global minima, resulting in sub-optimal solutions. However, when initialized near the optimum, they are highly effective \cite{gd-1, gd-2, dc3}. Therefore, in this paper, we propose randomly sampling a small number of points in the solution space to initialize the optimization process. 


 \begin{algorithm}[t]
 \caption{MicroOpt Algorithm}
  \label{algo}
 \begin{algorithmic}[1]
 \renewcommand{\algorithmicrequire}{\textbf{Input:}}
 \renewcommand{\algorithmicensure}{\textbf{Output:}}
 \REQUIRE Traffic $\boldsymbol{x}^s_{i}$, Slice Model $f^s_{\textit{QoS}}(\boldsymbol{x}_{i}^s, \boldsymbol{r}_{i}^s)$, QoS threshold $q^s_{\textit{thresh}}$, QoS degradation threshold $\beta^s_{\textit{thresh}}$, $\tau_{1, max}, \tau_{2, max}, \alpha_1, \alpha_2, \alpha_3, \epsilon_1, \epsilon_2$
 \ENSURE Optimal  resource allocation vector $\boldsymbol{r}^s_{i}$
  \STATE Initialize $\boldsymbol{\lambda}, \boldsymbol{\mu}, \textsc{LB}=0, \textsc{UB}=\infty, \tau_1=0, \tau_2=0$
  \vspace{3pt}
  \WHILE{ $\frac{\textsc{UB} - \textsc{LB}}{\textsc{UB}} > \epsilon_1$ \OR $\tau_1 < \tau_{1, max}$ }
    \STATE $\mathbf{r} \leftarrow$ \texttt{Initialization}($\boldsymbol{x}^s_{i}, f_{\textit{QoS}}(\boldsymbol{x}^s_{i}, \mathbf{r})$)
    \WHILE{ $|\nabla_r \mathcal{\hat{L}}| > \epsilon_2$ \OR $\tau_2 < \tau_{2, max}$ }
        \STATE $\mathbf{r}\leftarrow [\mathbf{r} - \alpha_1 \nabla_r \mathcal{\hat{L}}]^
+$
        \STATE $\tau_2 \leftarrow \tau_2 + 1$
    \ENDWHILE
    \STATE $\lambda_s \leftarrow [\lambda_s + \alpha_2(\beta_i^s - \beta^s_{\textit{thresh}})]^+,\,\forall s$
    \STATE $\mu_k \leftarrow [\mu_k + \alpha_3(\sum_{s \in \mathbf{S}} r^{s, k} - R^k)]^+,\,\forall k$
    \STATE $\textsc{LB} = \max(\textsc{LB}, \mathcal{L}(\mathbf{r}, \boldsymbol{\mu}, \boldsymbol{\lambda}))$
    \STATE $\textsc{UB} = \min(\textsc{UB}, \sum_{s \in {S}} \boldsymbol{\eta}^\intercal  \mathbf{r}^s)$
    \STATE $\tau_1 \leftarrow \tau_1 + 1$
  \ENDWHILE
 \RETURN $\mathbf{r}$ 
 \end{algorithmic} 
 \end{algorithm}

Algorithm~\ref{algo} outlines the steps of our proposed solution. The algorithm takes as input the traffic $\boldsymbol{x}^s_{i}$ and the model $f^s_{\textit{QoS}}(\boldsymbol{x}_{i}^s, \mathbf{r}_{i}^s)$ for each slice for the duration of reconfiguration interval. It also requires the QoS threshold $q^s_{\textit{thresh}}$, the QoS degradation threshold $\beta^s_{\textit{thresh}}$ specific to each slice, and several hyper-parameters to control the algorithm's behaviour. These include parameters related to the stopping condition, such as $\tau_{1, max}$ and $\tau_{2, max}$, which define the maximum number of iterations for the outer and inner loops, respectively, and $\epsilon_1$ and $\epsilon_2$ that determine the desired level of convergence for the upper and lower bounds of the objective function. We also have learning rates, $\alpha_1$, $\alpha_2$, and $\alpha_3$, for updating resource allocations and Lagrangian multipliers. Finally, the output of the algorithm is the optimal resource allocation for each slice.

The algorithm is comprised of outer and inner loops. Within the inner, the resource allocation variables are first initialized using through random sampling. Subsequently, these variables are updated using the gradient of the surrogate Lagrangian function ($\nabla_r \mathcal{\hat{L}}$). These updated variables are then projected into the non-negative domain, denoted by the notation $[.]^+$. After updating the resource allocation variables, the algorithm updates the Lagrange multipliers inside the outer loop. QoS constraints multipliers, $\lambda_s$, are updated based on the QoS degradation values $\beta^s_{i}$ and threshold $\beta^s_{\textit{thresh}}$ for each slice. Similarly, resource constraints multipliers, $\mu_k$, are updated for each resource $k$ by considering the difference between the allocated resources $\sum_{s \in {S}} r_{i}^{s, k}$ and the resource capacity $R^k$.
At each point, the upper bound $\textsc{UB}$  is equal to the best feasible solution found so far, while the lower bound $\textsc{LB}$ is equal to the value of the Lagrangian function. Once the termination condition is met, the algorithm returns the resource allocation corresponding to the best $\textsc{LB}$. 

%% file: implementation.tex
\section{Implementation}\label{sec:implementation}
\begin{figure}[ht!]
    \centering
    \includegraphics[width=\linewidth]{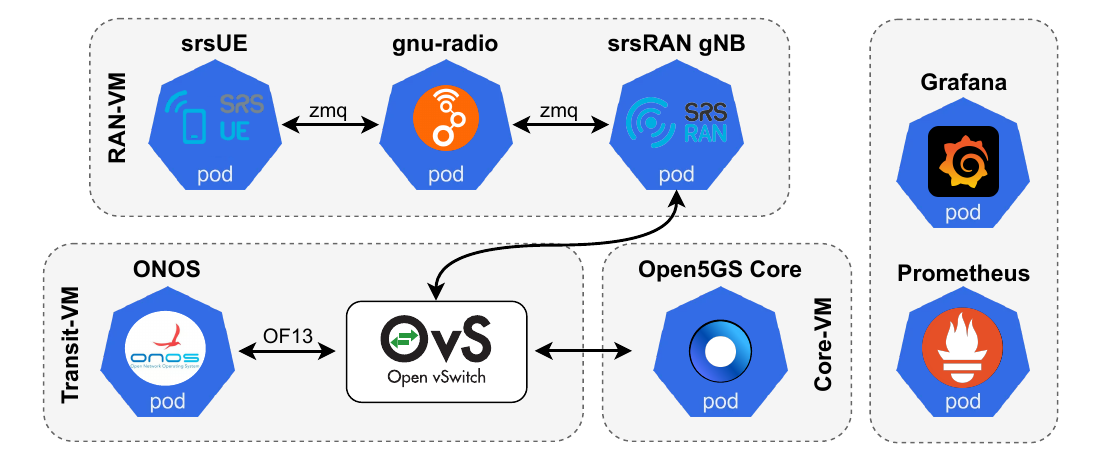}
    \caption{Overview of our 5G testbed}
    \label{fig:testbed}
\end{figure}

In this section, we describe the implementation of our network slicing testbed, shown in Fig.~\ref{fig:testbed}. 


\subsection{Testbed Infrastructure} \label{sec:testbed}


Our testbed consists of a substrate network deployed on a three-node Azure cluster. The virtual machines hosting the RAN, core, and transit networks are allocated 32 vCPUs, 16 vCPUs, and 8 vCPUs, respectively. Additionally, these virtual machines have RAM allocations of 64 GB, 32 GB, and 32 GB, respectively. This physical topology forms the foundation for our testbed, providing necessary computational and networking resources to support various network functions.

\subsection{5G Network Implementation}

\textbf{RAN.} We implement the 5G RAN using the srsRAN project \cite{srsran}, an open-source software designed to create a 3GPP Release 17 (R17) compliant gNB. The User Equipments (UEs) are implemented using srsUE \cite{srsUE}, a software implementation of a UE. Instead of physical radios for over-the-air transmissions between the gNB and UEs, we use virtual radios, also provided by srsRAN. Additionally, we utilize GNU Radio Companion to manage uplink and downlink signal between the UE and the gNB. GNU Radio offers a variety of signal processing blocks, enabling the emulation of complex functions such as path loss and user mobility.

\smallskip
\noindent \textbf{Core.} We implement the 5G mobile core based on Open5GS \cite{open5gs}, an open-source implementation of 3GPP R17. The network functions, \eg AMF, SMF, UPF and NRF are containerized and deployed on our Kubernetes cluster. We consider a network slicing scenario where each slice has dedicated UPF and SMF network functions, while sharing other common 5G core functions such as AMF and NRF.

\smallskip
\noindent \textbf{Transport.} To establish the transport network infrastructure, we employ a software-defined VXLAN overlay utilizing Open vSwitch (OvS) \cite{ovs} on the underlying physical substrate network. The integration of the 5G network functions with this transport overlay was accomplished by utilizing the OvS CNI plugin for Kubernetes. 


\subsection{Control and Management}
\noindent \textbf{MANO.} We use Kubernetes v1.29 for the management and orchestration (MANO) of our 5G network. This allows us to encapsulate different 5G network functions, including the RAN, into lightweight and portable containers. These containers can be dynamically deployed, scaled, and managed across our distributed cluster of nodes, providing flexibility and scalability. The Kubernetes API allows us to control the placement of these functions, and create network slices with desired topologies. To dynamically scale the CPU resources allocated to the network functions, we use Linux cgroups.
\smallskip

\noindent \textbf{SDN Controller.} We use the ONOS SDN controller \cite{onos} to enable precise control over the routing of network flows within our network slices. By communicating with the OvS switches in the VXLAN transport overlay, ONOS facilitates the routing of network slice traffic through OvS queues with specific rates, thus providing bandwidth slicing capabilities.

\subsection{Data Collection} \label{sec:dataset}

\noindent \textbf{Edge Application.} To test the proposed solution, we focus on generating enhanced Mobile Broadband (eMBB) user traffic. Note that we previously defined \textit{slice traffic} as the number of users connected to a slice (users/s), whereas \textit{user traffic} defined here is the traffic generated by the UEs (Mbps). To generate the user traffic, we begin by collecting a packet capture (pcap) dataset for 4K video streaming from YouTube. This dataset is then replayed whenever a UE connects to the eMBB slice. Streaming 4K video can consume significant network resources, potentially degrading the QoS for other slices within the network. Therefore, it is crucial to dynamically scale the bandwidth allocation in the transport network and adjust the gNB CPU allocation for the eMBB slice.

\smallskip
\noindent \textbf{Monitoring.} To comprehensively monitor our testbed and gather datasets for our slice model, we implement a robust architecture using Prometheus and Grafana, as suggested by existing literature in  \cite{saha2023monarch}. Prometheus collects and stores time-series data on network traffic, resource utilization, and network function performance. Grafana provides a user-friendly platform for visualizing and analyzing this data.
\smallskip

\noindent \textbf{Dataset Collection.}
We define QoS as the end-to-end throughput received by the UE, with varying QoS thresholds ($q_{\textit{thresh}}$) of $3$-$5$ Mbps and acceptable QoS degradation thresholds ($\beta_{\textit{thresh}}$) of $0.1$-$0.3$. Note that, for brevity, we omit the index $s$ identifying the slice in the previously introduced notation for the rest of the paper. The targeted resources for scaling in this scenario are the transport network bandwidth and the gNB CPU with equal priority, \ie $\boldsymbol{\eta} = \left[\frac{1}{50}; \frac{1}{4500}\right]$, where $50$ Mbps and $4500$ millicores are the max resource allocations for the eMBB slice. Note that for the evaluation, the results show the normalized resource allocations, \ie $\boldsymbol{\eta}^\intercal \mathbf{r}_{i}$.

 To train the slice model, we gather a QoS dataset with the slice traffic varying from $1$ to $5$ users/s, the transport bandwidth from 5 to 40 Mbps, and the gNB CPU resource from 500 to 4000 millicores, in intervals of 1 users/s, 5 Mbps, and 500 millicores, respectively. 

%% file: experiments.tex
\section{Performance Evaluation}\label{sec:experiments}
\subsection{Experiment Setup and Comparison Approaches}
\subsubsection{Slice Model}

\begin{figure*}[ht!]
\captionsetup{justification=centering}
  \begin{minipage}{0.64\columnwidth}
  \centering
  \includegraphics[width=\linewidth]{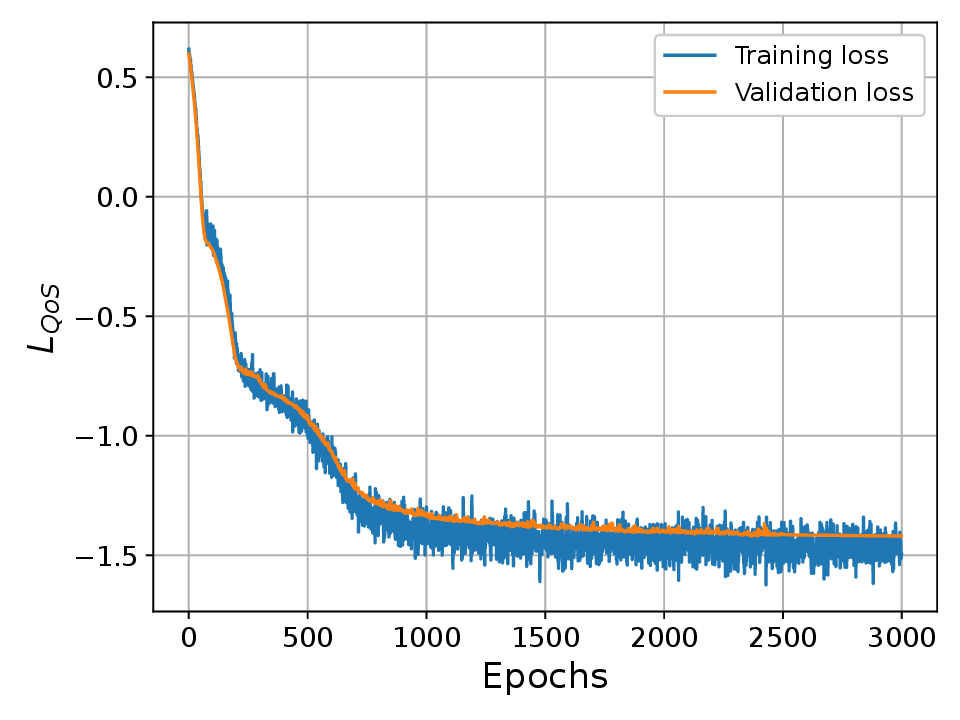}
    \subcaption{Negative Log probability loss ($L_{\textit{QoS}}$ in \eqref{eq:lqos});\,\, Test $L_{\textit{QoS}}$: $-1.67$}
    \label{fig:qos_loss}
  \end{minipage}
  \begin{minipage}{0.64\columnwidth}
  \centering
  \includegraphics[width=\linewidth]{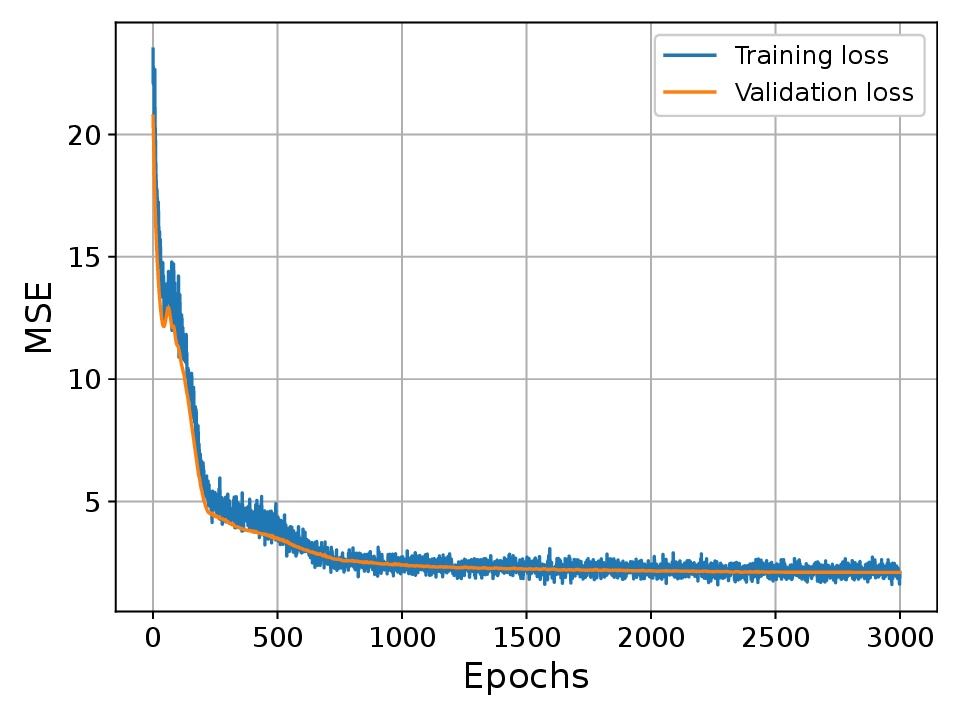}
    \subcaption{Mean Squared Error (MSE)\\ Test MSE: $0.62$}
    \label{fig:mse_loss}
  \end{minipage}
  \label{fig:model_training}
\begin{minipage}{0.64\columnwidth}
  \centering
  \includegraphics[width=\linewidth,trim={0.5cm -0cm 1cm 1.2cm},clip]{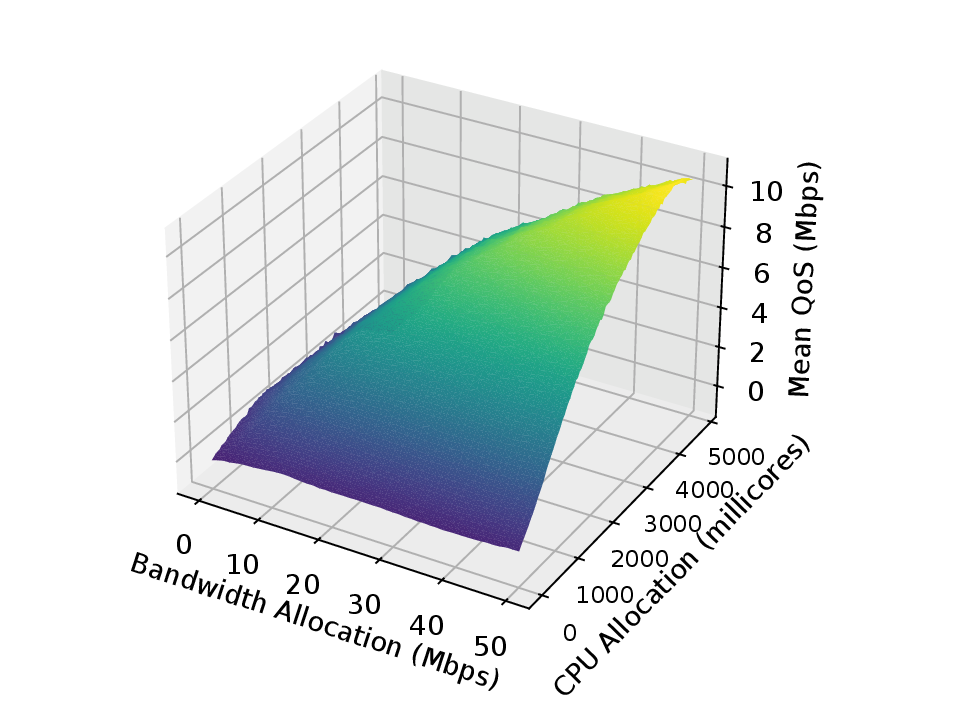}
    \subcaption{Slice model visualization}
    \label{fig:3d_model}
  \vspace{10pt}
  \end{minipage}
  \caption{Slice model training and visualization}
\end{figure*}


We presented a high-level overview of the slice model in \fig{fig:net_model}. However, the specific details of the model, including the number of layers, activation functions, and nodes per layer, may vary for different datasets collected from other testbeds. In our model, after the input layer, we add a  batch normalization layer to normalize the inputs which leads to improved model performance by reducing the internal covariate shift. Subsequently, we use three shared hidden layers with 16 nodes each and Rectified Linear Unit (ReLU) activation. The mean and standard deviation branches have one hidden layer with 16 nodes and ReLU activation.
The mean output uses a Linear activation function, while the standard deviation output employs the Softplus activation function to ensure non-negativity. It should be noted that the slice model’s specific details, such as its structure, the number of layers, activation functions, and nodes per layer, can vary for different testbeds.

We divide the QoS dataset (\cf \sect{sec:dataset}) into training and validation sets. Additionally, we perform the procedure outlined in \sect{sec:dataset} to gather a test set consisting entirely of off-grid points, \ie input combinations not present in the training or the validation sets. Subsequently, we train the model for 3,000 epochs with a learning rate of 0.001, and reduce the learning rate by a factor of 10 after 1,500 epochs. Figures \ref{fig:qos_loss} and \ref{fig:mse_loss} show the negative log probability loss (\ie $L_{QoS}$ in \eqref{eq:lqos}), and mean squared error (MSE), respectively, as the model trains. We can see that the MSE follows the same trend as $L_{QoS}$. Additionally, from the figures, we can see that the validation error does not deviate from the training error, which shows that the model is not overfitting to the training data. Once trained, the slice model achieves a $L_{QoS}$ of -1.53, -1.72, and -1.67, an MSE of 1.58, 1.24, and 0.91, and a mean absolute error (MAE) of 0.69, 0.62 and 0.54 on the training, validation, and test datasets, respectively.

Finally, we visualize the trained slice model by plotting the mean QoS predicted at the entire range of input values in Fig.~\ref{fig:3d_model}. This figure shows the predicted QoS mean, averaged over 1-5 users/s. We can see that at lower CPU allocations, the QoS does not increase with bandwidth, as the gNB is the bottleneck. Once the CPU allocation exceeds $100$ millicores, the effect of bandwidth on the QoS becomes significant. 

\subsubsection{Comparison Approaches}\label{sec:comparison_approaches}
To evaluate the proposed approach, we implement the following state-of-the-art approaches:

\smallskip
\noindent \textbf{Peak-alloc}: This approach serves as a baseline where the network operator statically allocates resources to a slice based solely on peak slice traffic (5 users/s) and strict QoS requirements ($q_{\textit{thresh}}=5.0$, $\beta_{\textit{thresh}}=0.01$). To determine the optimal resource allocation for this scenario, we employ a brute-force method, specifically a fine-grained grid search, which took several hours to find the solution.

\smallskip
\noindent \textbf{Atlas, Altas$^+$}: This approach, based on \cite{liu2022atlas}, uses a network simulator to train a Bayesian neural network (BNN) for learning the QoS degradation function in eqn.~\eqref{beta}. BO with Thompson sampling and Lagrangian relaxation is then used for resource optimization. In our implementation, we replace the network simulator with our slice model. The original method in \cite{liu2022atlas} works only for constant slice traffic (\ie peak traffic in the slice traffic distribution) because it requires pre-training the BNN with an expensive simulator, which is impractical for infinite possible traffic distributions. We propose using a DNN-based slice model, which makes the pre-training much faster and can be done after predicting the actual slice traffic distribution. This allows the method to handle slice traffic distributions. We refer to the original and the enhanced methods as Atlas and Atlas$^+$, respectively.

\smallskip
\noindent \textbf{CaDRL, CaDRL$^+$}: Several works have proposed constrained DRL for dynamic resource scaling \cite{noms23, liu2021constraint, icnp_older}. We implement the recent approach by \citet{icnp_older}, which uses Interior-point Policy Optimization (IPO) for this purpose, known as Constraint-aware Deep Reinforcement Learning (CaDRL). However, constrained DRL approaches converge to the worst-case scenario (\ie peak traffic) when trained to generalize over multiple slice traffic distributions, as shown by \citet{noms23}. We propose using a computationally inexpensive slice model to quickly train the policy from scratch once the actual slice traffic has been predicted, avoiding the need to generalize over multiple distributions. We refer to these methods as CaDRL and CaDRL$^+$, respectively.

\subsubsection{Slice Traffic Model} \label{sec:traffic}

\begin{figure}[!t]
	\centering
        \includegraphics[width=0.75\linewidth]{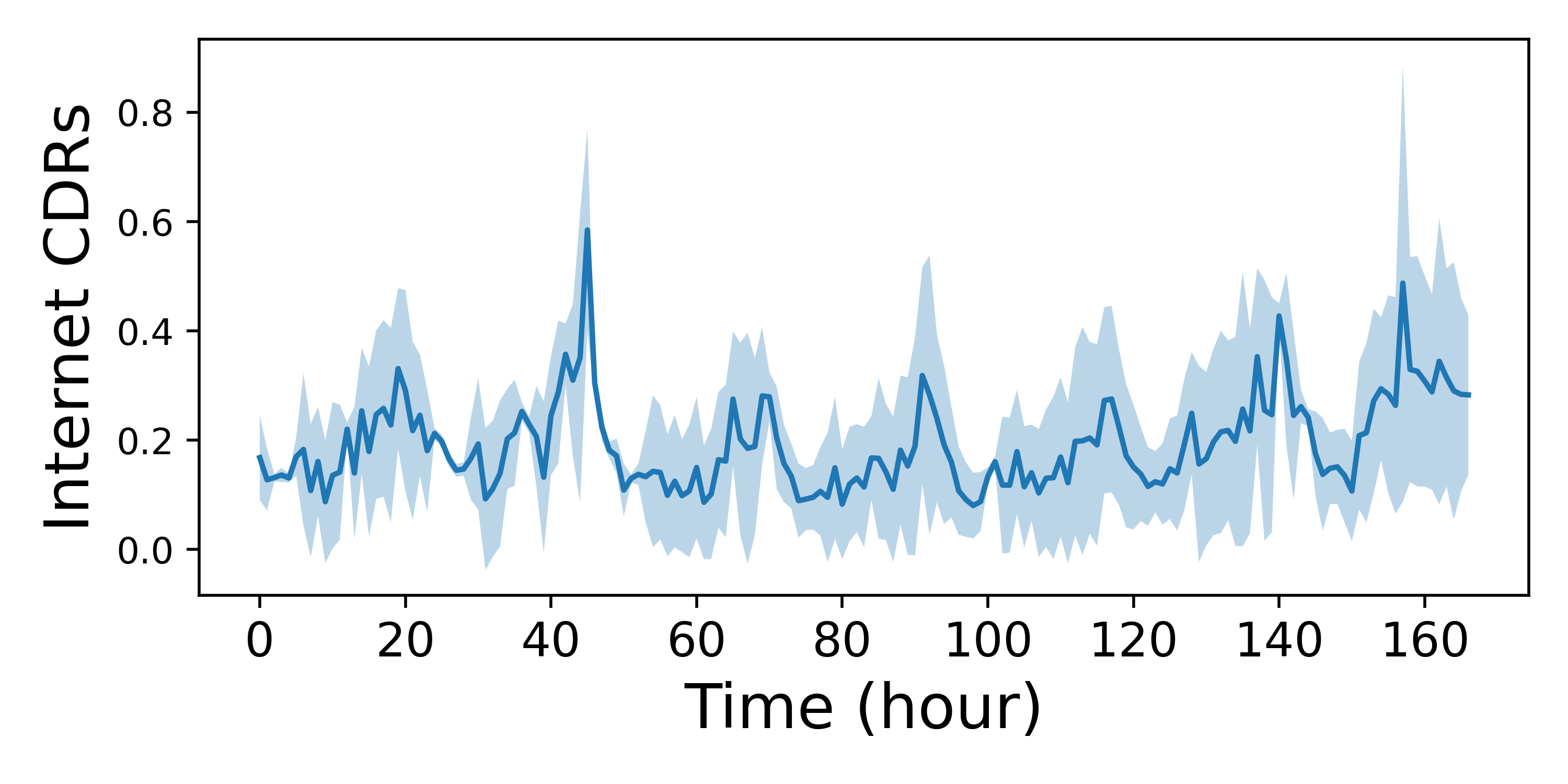}
        \caption{Telecom Italia dataset traffic trend over a week}
        \label{dataset_traffic}
\end{figure}

To generate different slice traffic distributions, we use the Telecom Italia dataset \cite{traffic_dataset}, which contains anonymized telecommunication activity in Milan and the Province of Trentino. Focusing on an eMBB slice, we extract one week of internet call detail records (CDRs), which are generated every time a user initiates or ends an internet connection. \fig{dataset_traffic} illustrates the normalized traffic variation for a randomly selected cell, showing significant hourly variation. Subsequently, we calculate the mean and max standard deviation of this data, referred to as $\sigma_\textit{mean}$ and $\sigma_\textit{max}$, respectively. We then generate 10 different slice traffic distributions using a truncated normal distribution centered at 1-5 users/s with standard deviations of $\sigma_\textit{mean}$ and $\sigma_\textit{max}$,  referred to as dataset slice traffic distributions. Once a user connects to the slice, the user traffic is generated as described in Section \ref{sec:dataset}.

\begin{figure*}[t!]
\centering
\begin{minipage}{0.66\columnwidth}
  \centering
  \includegraphics[width=\linewidth]{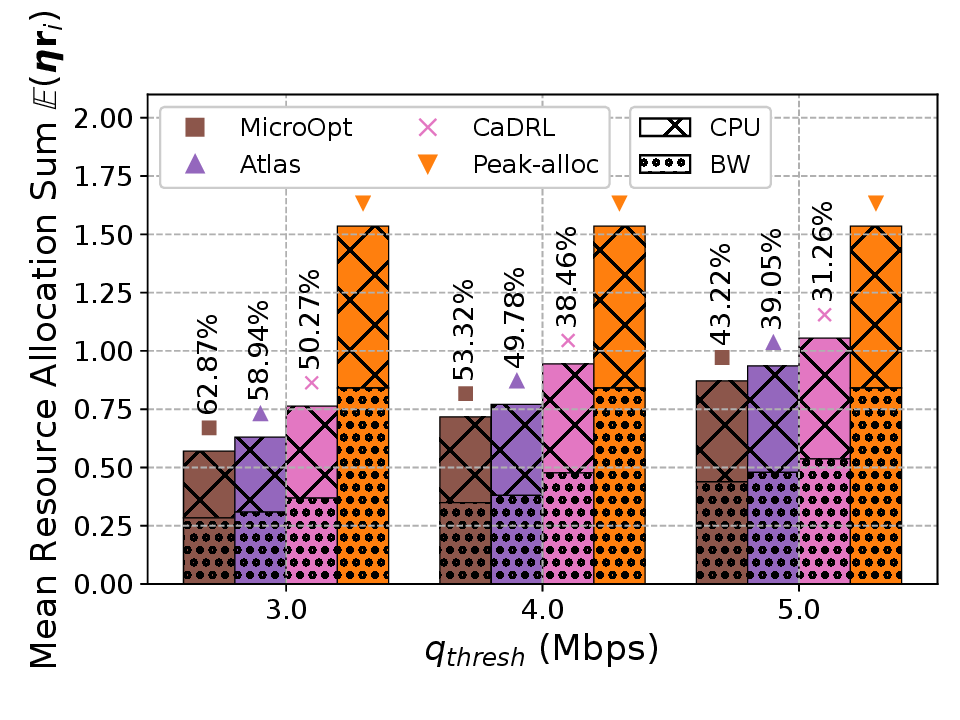}
    \subcaption{Mean resource allocation vs. QoS threshold ($q_{\textit{thresh}}$), averaged QoS degradation threshold ($\beta_{\textit{thresh}}$)}
    \label{fig:qos_trend}
  \end{minipage}
  \hfill
  \begin{minipage}{0.66\columnwidth}
  \centering
  \includegraphics[width=\linewidth]{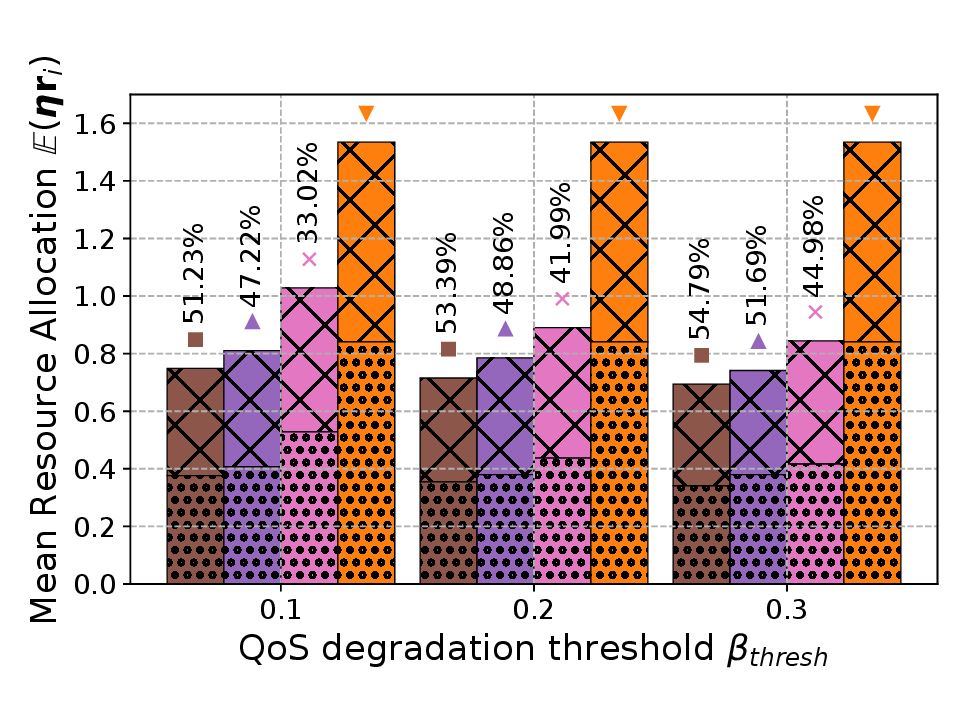}
    \subcaption{Mean resource allocation vs. QoS degradation threshold ($\beta_{\textit{thresh}}$), averaged over QoS threshold ($q_{\textit{thresh}}$)}
    \label{fig:satis_trend}
  \end{minipage}
  \hfill
  \begin{minipage}{0.66\columnwidth}
  \centering
  \includegraphics[width=\linewidth]{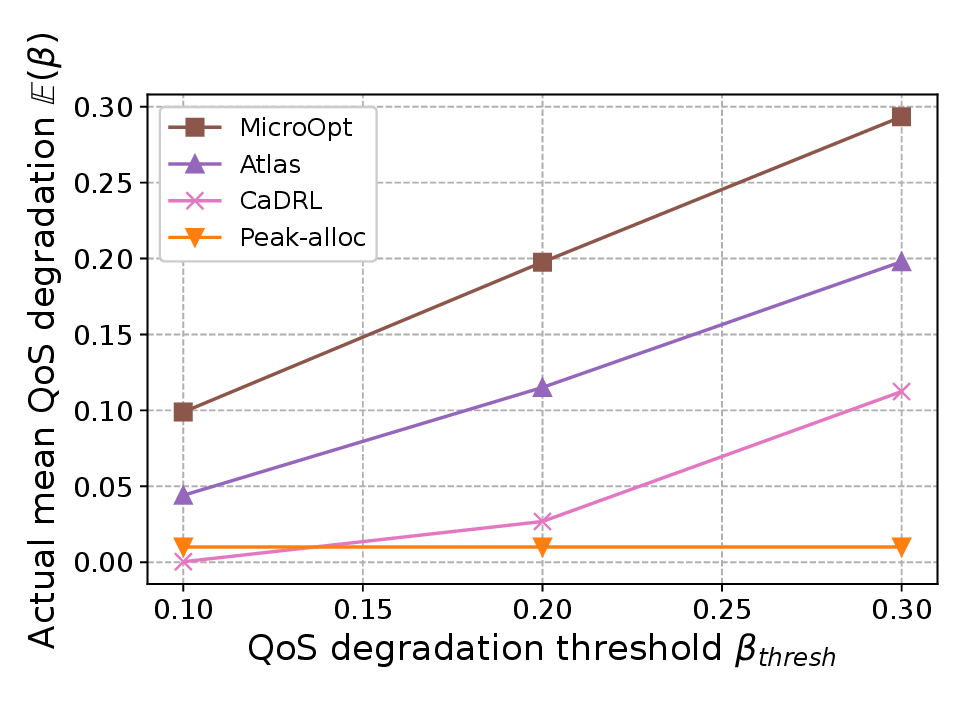}
    \subcaption{Mean QoS degradation $\mathbbm{E}(\beta)$ vs. varying QoS degradation threshold $\beta_{\textit{thresh}}$, averaged over QoS threshold ($q_{\textit{thresh}}$)}
    \label{fig:deg_trend}
  \end{minipage}
  \caption{Mean resource allocation sum $\mathbb{E}(\boldsymbol{\eta} \mathbf{r}_{i})$ and mean QoS degradation $\mathbb{E}(\beta)$ at various parameter settings}
  \label{fig:vary_res_alloc}
\end{figure*}

\begin{figure*}[ht!]
\centering
\begin{minipage}{0.66\columnwidth}
  \centering
  \includegraphics[width=\linewidth]{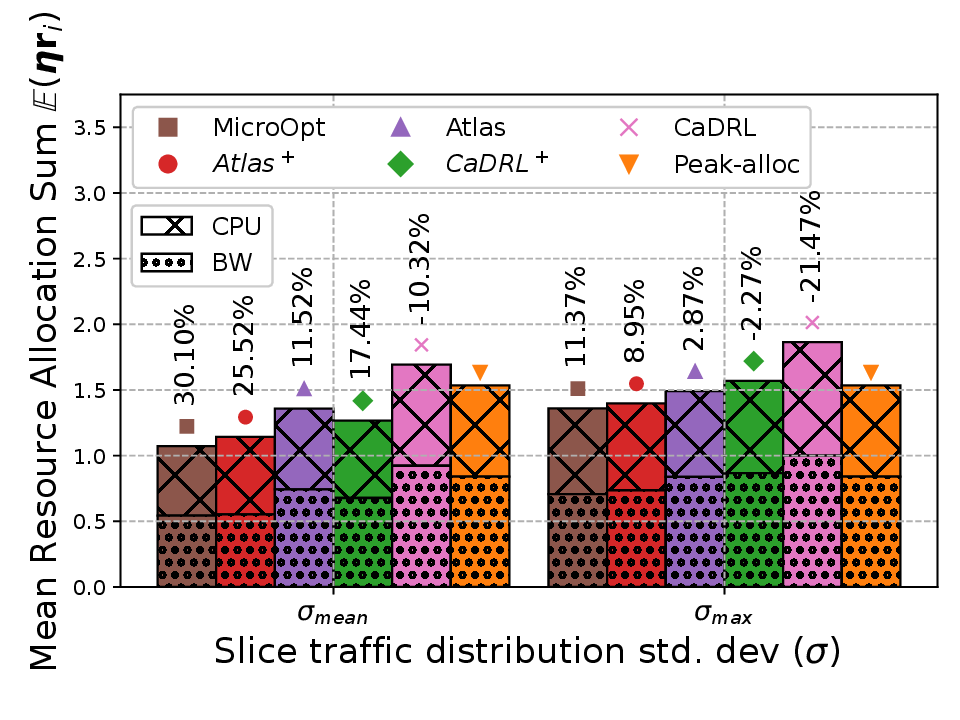}
    \caption{Mean resource allocation vs. slice traffic std. dev.}
    \label{fig:user_dist_res_alloc}
  \end{minipage}
  \hfill
  \begin{minipage}{0.66\columnwidth}
  \centering
  \includegraphics[width=\linewidth]{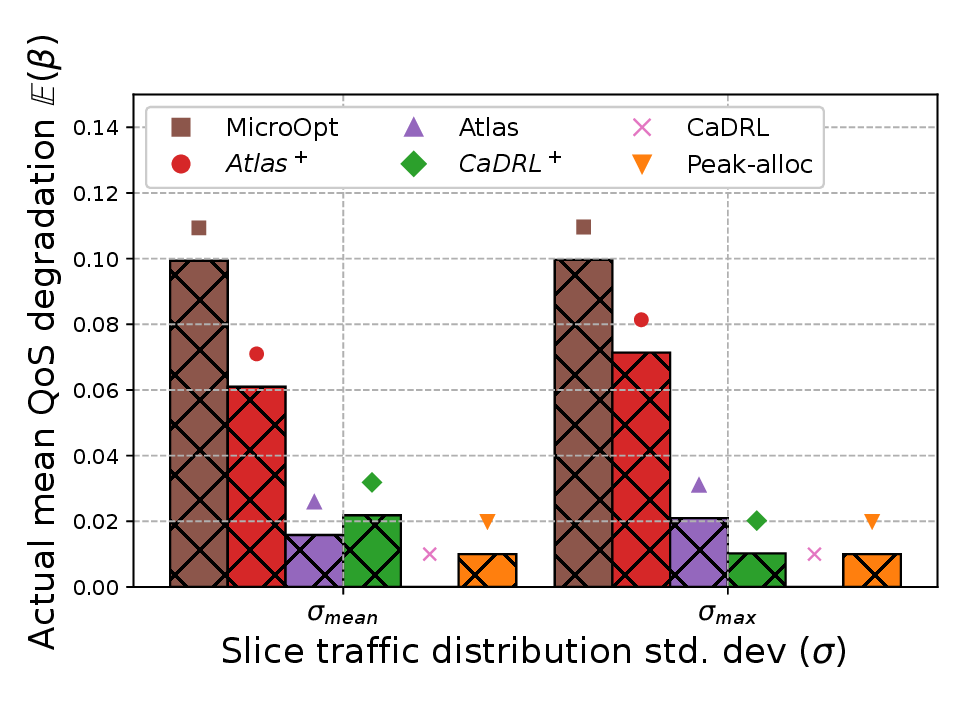}
    \caption{Mean QoS degradation $\mathbb{E}(\beta)$ vs. slice traffic std. dev.}
    \label{fig:user_dist_deg}
  \end{minipage}
  \hfill
\begin{minipage}{0.66\columnwidth}
  \centering
  \includegraphics[width=\linewidth]{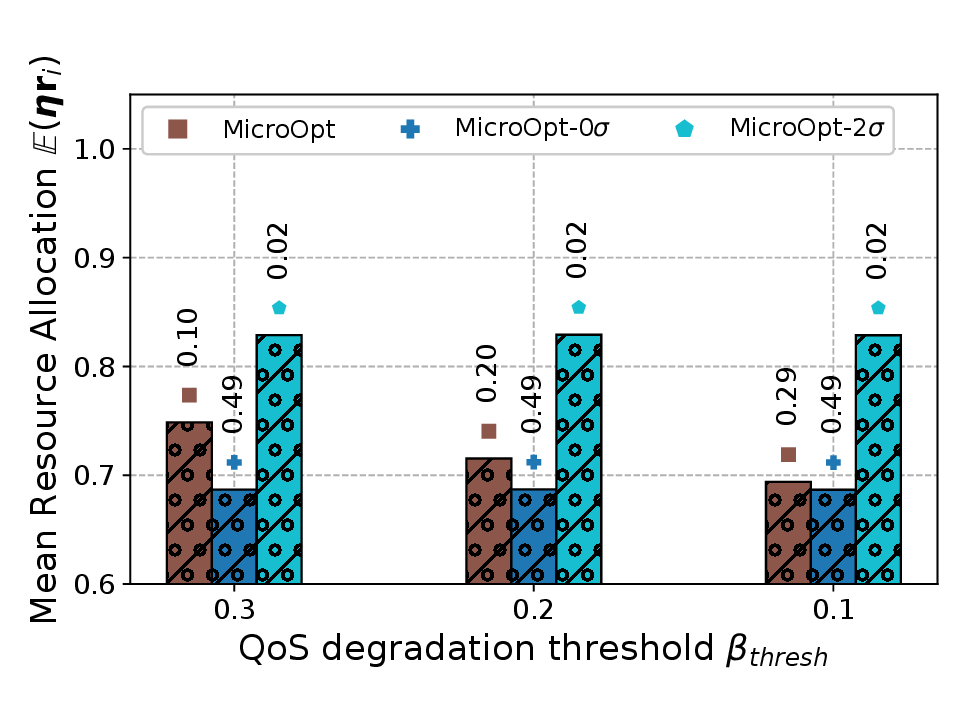}
    \caption{Mean QoS degradation ($\mathbbm{E}(\beta)$) considering scalar vs. distributed QoS}
    \label{fig:ablation}
  \end{minipage}
\end{figure*}

\subsection{Simulation Results} \label{sec:sim_results}

In this section, we compare different approaches for deriving the optimal resource allocation of the constrained optimization problem in \eqref{obj}, given the network model (\cf \sect{sec:nn_model}), and the slice traffic $\boldsymbol{x}_{i}$ (\cf \sect{sec:traffic}).

\begin{table}[htbp]
\footnotesize
  \centering
  \caption{Resource allocation and QoS degradation}
  \label{tab:data}
  \begin{tabular}{lcccc}
    \toprule
    Approach & \makecell{Slice traffic ($\boldsymbol{x}_{i}$) \\ (users/s)} & \makecell{CPU \\ (millicores)} & \makecell{BW \\ (Mbps)} & \makecell{QoS degr. \\ $\mathbbm{E}(\beta)$} \\
    \midrule
    \multirow{5}{*}{Atlas} & 1 & 1167.60 & 10.03 & 0.06 \\
                           & 2 & 1648.72 & 14.84 & 0.07 \\
                           & 3 & 2452.79 & 29.79 & 0.00 \\
                           & 4 & 2589.30 & 30.89 & 0.05 \\
                           & 5 & 2934.51 & 42.11 & 0.07 \\
    \midrule
    \multirow{5}{*}{CaDRL} & 1 & 1632.14 & 13.27 & 0.00 \\
                           & 2 & 2011.99 & 19.79 & 0.00 \\
                           & 3 & 2574.88 & 33.23 & 0.00 \\
                           & 4 & 3089.35 & 47.89 & 0.00 \\
                           & 5 & 3890.13 & 50.00 & 0.00 \\
    \midrule
    \multirow{5}{*}{MicroOpt} & 1 & 993.93 & 11.73 & 0.10 \\
                              & 2 & 1463.44 & 15.77 & 0.10 \\
                              & 3 & 2140.87 & 22.22 & 0.10 \\
                              & 4 & 2534.28 & 28.94 & 0.10 \\
                              & 5 & 2945.72 & 39.20 & 0.10 \\
    \midrule
    Peak-alloc & 5 & 3123.80 & 42.20 & 0.01 \\
    \bottomrule
  \end{tabular}
\end{table}

\smallskip
\noindent \textbf{Constant Slice Traffic.} First, we evaluate the different approaches at a constant slice traffic varying from 1-5 users/s, with a QoS threshold of $q_{\textit{thresh}} = 5.0$ and a QoS degradation threshold of $\beta_{\textit{thresh}} = 0.1$.  Atlas$^+$ and CaDRL$^+$ are excluded from this evaluation because, with a constant number of users, they perform the same as their counterparts. Table \ref{tab:data} presents the corresponding resource allocation and QoS degradation achieved by the different approaches in this scenario. From the table, we can see that MicroOpt allocates the minimum resources while keeping the QoS degradation $\mathbbm{E}(\beta)$ under the required threshold of 0.1. CaDRL, despite achieving similar resource allocation as the other approaches, fails to increase the QoS degradation above 0, resulting in the highest resource allocation. Atlas maintains a QoS degradation close to 0.1 in most scenarios, with resource allocation similar to MicroOpt. Finally, Peak-alloc allocates more resources compared to MicroOpt and Atlas, as it only finds the optimal solution for a QoS degradation of $\beta_{\textit{thresh}} = 0.01$.

Next, we evaluate the different approaches across varying QoS and QoS degradation thresholds, omitting the tabular presentation of results for brevity. \fig{fig:qos_trend} and \fig{fig:satis_trend} show the normalized mean resource allocation and QoS degradation for various approaches across different values of $q_{\textit{thresh}} = [3.0, 4.0, 5.0]$ and $\beta_{\textit{thresh}} = [0.1, 0.2, 0.3]$, along with the percentage improvement over the baseline Peak-alloc. \fig{fig:qos_trend} shows that the proportions of CPU and bandwidth resources within the overall resource allocations are both approximately 1/2. This is because we used equal weights for CPU and bandwidth when solving \eqref{obj}, \ie with $\boldsymbol{\eta} = \boldsymbol{1} \cdot \left[\frac{1}{50}, \frac{1}{4500}\right]$.

From the figures, we can see a direct relationship between the QoS requirement ($q_{\textit{thresh}}$) and resource allocation---as the QoS requirement decreases, the resource allocation also decreases. In contrast, \fig{fig:satis_trend} demonstrates an inverse relationship between resource allocation and the QoS degradation threshold ($\beta_{\textit{thresh}}$)---as the QoS degradation threshold increases, indicating a tolerance for higher degradation, a lower resource allocation is required. Across different parameter values, the optimal resource allocation follows a consistent pattern---MicroOpt allocates the lowest resources, followed by Atlas, CaDRL, and Peak-alloc. Averaging over all parameter values, the respective approaches allocate 0.719, 0.778, 0.920, and 1.534 units of resources. This shows that MicroOpt results in a 7.65\%, 21.90\%, and 53.14\% decrease in resource usage compared to Atlas, CaDRL, and Peak-alloc, respectively.

\fig{fig:deg_trend} shows the QoS degradation achieved by the different approaches as they minimize resource allocation. From the figure, we see that all approaches achieve mean QoS degradation below the required threshold. Additionally, the mean QoS degradation follows the opposite trend to resource allocation---Peak-alloc achieves the lowest QoS degradation, followed by CaDRL, Atlas, and MicroOpt. 
Interestingly, unlike the previous tabular results, CaDRL achieves a higher QoS degradation and a lower resource allocation than Peak-alloc at more relaxed parameter settings (\ie lower $q_{\textit{thresh}}$, and higher $\beta_{\textit{thresh}}$). This is because CaDRL's algorithm highly prioritizes maintaining strict QoS degradation adherence over minimizing resource allocation.  Therefore, it is able to perform better when the QoS degradation threshold is lower.

\smallskip
\noindent \textbf{Dynamic Slice Traffic.} To evaluate the different approaches when slice traffic varies within a resource reconfiguration interval, \ie when slice traffic is a distribution rather than constant, we use the slice traffic distributions dataset  described in Section \ref{sec:traffic}. \fig{fig:user_dist_res_alloc} shows the mean resource allocation and the percentage improvement over Peak-alloc, while \fig{fig:user_dist_deg} illustrates the mean QoS degradation by the different approaches at a QoS threshold of $q_{\textit{thresh}} = 5.0$ and a QoS degradation threshold of $\beta_{\textit{thresh}} = 0.1$. 

From \fig{fig:user_dist_res_alloc}, it is evident that as the slice traffic std. dev. increases from $\sigma_{\textit{mean}}$ to $\sigma_{\textit{max}}$, the corresponding resource allocation also increases. This is because a higher std. dev. indicates a broader distribution of slice traffic, which necessitates more resources to maintain the mean QoS for the entire range of users, including those in the tail of the distribution.

Focusing on the resource allocation across the different approaches in \fig{fig:user_dist_res_alloc}, we can see that it follows the same trend as in the constant traffic scenario---MicroOpt allocates the least resources, compared to Atlas and CaDRL. However, with dynamic slice traffic, Atlas$^+$ and CaDRL$^+$ allocate fewer resources than their counterparts. This is because, as described in Section \ref{sec:comparison_approaches}, these approaches consider the actual slice traffic distribution rather than just the peak traffic. In this scenario, MicroOpt shows a mean decrease of 4.23\%, 14.23\%, 14.60\%, 31.61\%, and 20.74\% compared to Atlas$^+$, CaDRL$^+$, Atlas, CaDRL, and Peak-alloc, respectively, across the 10 different slice traffic distributions.

In \fig{fig:user_dist_deg}, we observe that all the different approaches achieve QoS degradation below the required threshold of $\beta_{\textit{thresh}} = 0.1$. Additionally, the QoS degradation follows the expected trend based on resource allocations, \ie approaches with higher resource allocation result in lower QoS degradation. In this scenario ($q_{\textit{thresh}} = 5.0, \beta_{\textit{thresh}} = 0.1$), RL-based approaches perform worse compared to the Peak-alloc approach. This is because RL-based approaches provide sub-optimal solutions, and when the parameters are similar to those of Peak-alloc, which is derived from a brute-force approach, the sub-optimality becomes evident.

\subsection{Ablation Study}
The slice model proposed in Section \ref{sec:nn_model} learns a QoS distribution for any given input, requiring the use of the reparameterization trick for gradient calculation. The output QoS can be a distribution due to factors such as user mobility, non-deterministic transmission medium, and network function behavior. In this section, we investigate the scenario where the slice model only learns a scalar QoS value, \ie the mean ($\mu$) or two std. dev. below the mean (\ie $\mu - 2\sigma$) in order to ensure QoS satisfaction. This allows direct gradient computation but ignores the actual underlying distribution. We refer to these approaches as MicroOpt-0$\sigma$ and MicroOpt-2$\sigma$, respectively.

For evaluation, we solve the constrained optimization for constant slice traffic under different QoS and QoS degradation thresholds (\cf Section \ref{sec:sim_results}). \fig{fig:ablation} shows the mean resource allocation and the corresponding QoS degradation achieved at various QoS degradation thresholds. From the figure, we observe that although MicroOpt-0$\sigma$ allocates the least resources, it maintains a QoS degradation of 0.49 which is significantly higher than the thresholds. This occurs because, with a normal distribution, half of the QoS values fall below the mean, leading to a QoS degradation of 0.5. On the other hand, MicroOpt-2$\sigma$ achieves small QoS degradation but allocates significantly higher resource than MicroOpt. Finally MicroOpt considers the QoS as a distribution, accounting for deviations from the mean or tail QoS. This leads to the highest resource saving while maintaining the QoS degradation under the required threshold. This ablation study highlights the importance of modeling QoS as a distribution.

\subsection{Testbed Evaluation}

\begin{figure}
  \centering
  \includegraphics[width=0.6\linewidth]{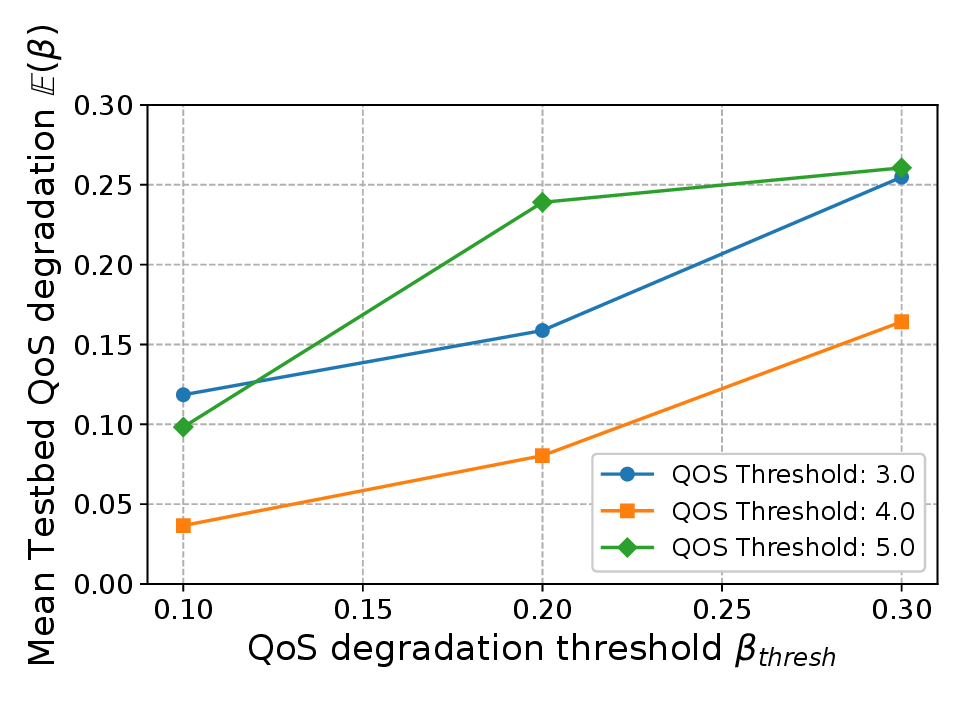}
    \caption{Testbed Mean QoS degradation ($\mathbbm{E}(\beta)$) of MicroOpt}
    \label{fig:testbed_eval}
\end{figure}

 The feasibility of solutions determined by the slice model may not necessarily translate to feasibility within the real-world network if the model is inaccurate. Therefore, even though the proposed approach achieves the least resource consumption while satisfying QoS degradation constraints (\cf \fig{fig:deg_trend}) based on the slice model, these solutions must be validated on the testbed to ensure the accuracy of the slice model. For this purpose, we generate the slice traffic with $\sigma_{mean}$ described in Section \ref{sec:traffic} on our testbed by using automated scripts for the arrival and departure of UEs. The resource allocation is then determined by solving the constrained optimization problem using Algorithm~\ref{algo}.
  

\fig{fig:testbed_eval} shows the mean QoS degradation ($\mathbbm{E}(\beta)$) of MicroOpt on the 5G testbed across various QoS thresholds ($q_{\textit{thresh}}$) and QoS degradation thresholds ($\beta_{\textit{thresh}}$). The figure indicates a positive correlation between the QoS degradation threshold and the mean QoS degradation. Most parameter satisfy the QoS degradation threshold However, for specific configurations ($\beta_{\textit{thresh}}, q_{\textit{thresh}}=(0.1, 3.0)$ and $(0.2, 5.0)$), it slightly exceeds the threshold. This occurs when the slice model overestimates QoS, leading to higher than expected QoS degradation. This limitation is inherent to approaches using surrogate models of the real network and can be mitigated by introducing a safety margin, fine-tuning the solution during online operation \cite{liu2022atlas}, or switching to a safe solution if the QoS degradation exceeds the threshold \cite{liu2021onslicing}.

To gauge the extent of our slice model's under-prediction, we evaluated the different losses of the  model using a newly gathered dataset that includes evaluation slice traffic, the resource allocations performed, and the corresponding QoS achieved. On this data, the model exhibited a $L_{QoS}$ of -1.53, an MSE of 1.14, and an MAE of 0.72. These values are consistent with the training, validation, and test losses, indicating the model's reliability. The alignment of these metrics across different datasets suggests that as the slice model undergoes further fine-tuning, it can more accurately predict QoS degradation. Consequently, the 5G testbed's QoS degradation will better align with the set thresholds.

%% file: conclusion.tex
\section{Conclusion}\label{sec:conclusion}
In this paper, we presented the MicroOpt framework, a novel approach for end-to-end dynamic resource allocation in 5G and beyond network slices. The framework leverages a DNN with the reparameterization trick to learn a differentiable slice model, which is then used in a primal-dual optimization algorithm to minimize the resource allocation under QoS constraints. We evaluated the proposed framework in multiple scenarios and showed that it can achieve up to 21.9\% reduction in resource allocation compared to the state-of-the-art approaches, while also satisfying the QoS degradation constraints. Finally, we deployed an open-source 5G testbed with data collection and scaling capability for validating the results and showed that the proposed solution is feasible in various scenarios.


Our future work will explore incorporating a feedback mechanism to address inaccuracies in traffic prediction or the slice model. Additionally, investigating resource allocation in scenarios with multiple slices and resource contention is essential for understanding how the MicroOpt framework handles competition for scarce resources.